# Antiferromagnetic order and consequences on the transport properties of $Ba_4Ru_3O_{10}$

Y. Klein[1*], G. Rousse[1], F. Damay[2], F. Porcher[2], G. André[2], and I. Terasaki[3]

[1]*Université Pierre et Marie Curie - Paris 6, IMPMC - CNRS UMR7590*
*Campus Jussieu, 4 place Jussieu 75252 Paris Cedex 05, France*
[2]*Laboratoire Léon Brillouin, CEA-CNRS UMR 12, 91191 Gif-sur-Yvette Cedex, France*
[3]*Department of Physics, Nagoya University, Nagoya 464-8602, Japan*

*Electronic address: yannick.klein@impmc.upmc.fr

**Abstract**

Barium ruthenate $Ba_4Ru_3O_{10}$, in which $Ru_3O_{12}$ trimers are connected together to form a chequered two-dimensional framework, has been synthesised and its structural, magnetic and transport properties studied between 300 K and 2 K. The paramagnetic to antiferromagnetic transition at $T_N \approx 105$ K evidenced on the susceptibility curve coincides with an increase of electron localization in transport measurements. Thermoelectric power and Hall coefficient measurements both exhibit dramatic changes at $T_N$, characteristic of a reconstruction of the bands structure near the Fermi level. No pronounced structural changes are observed at $T_N$ in this compound. The magnetic scattering signal on the neutron powder diffraction patterns below $T_N$ is weak, but can be tentatively modelled with an antiferromagnetic ordering of the spins at both ends of a trimer, the spin of the more symmetric Ru site remaining idle. Crystal field and strong spin-orbit coupling at the $Ru^{4+}$ site seem to be the key parameters to understand the magnetic state of $Ba_4Ru_3O_{10}$.

**Introduction**

Interplay between orbital, spin, charge carriers and lattice degrees of freedom has been a fascinating subject for the condensed-matter physics community for the last decades. In $3d$ transition metal oxides such as doped manganese perovskites [1] it can lead to a variety of charge and magnetic ground states, with properties as diverse as colossal magnetoresistance [2] or charge and orbital ordering [3]. Although $4d$ and $5d$ orbitals are more extended in space and hybridize with O $2p$ orbitals, which should result in itinerant states, strong spin-orbit coupling has been shown to induce a Mott instability even in the weakly correlated electron $5d$ system $Sr_2IrO_4$ [4]. Amongst $4d$ transition metal oxides, orbital ordering has been argued to result in the spontaneous formation of Haldane chains in the three-dimensional cubic ruthenium oxide $Tl_2Ru_2O_7$ [5] ; in $La_4Ru_2O_{10}$ the semiconductor-semiconductor transition, associated with a drop in the susceptibility as temperature decreases, has also been explained through a complex coupling between orbital, charge and spin degrees of freedom, involving a spin-ladder-like spin-singlet dimerization [6-10].

Although cubic or layered-perovskite ruthenates have been largely investigated, more unusual structural topologies still remain vastly understudied. Amongst them, $Ba_4Ru_3O_{10}$ [11] exhibits an orthorhombic *Cmca* structure [12], in which $Ru_3O_{12}$ trimers of face-shared $RuO_6$ octahedra are connected together via their corners, to build corrugated layers that are then stacked along the *b*-axis (Figure 1). Ru atoms are distributed on two symmetrically unequivalent sites : Ru(1), at the centre of the trimer, is on Wyckoff position $4a$, and the two outer Ru(2) atoms are on position $8f$. The first magnetic susceptibility ($\chi$) measurements carried out on $Ba_4Ru_3O_{10}$ [11] showed a Curie-Weiss behaviour at high temperature, followed at $T$ = 105 K by an abrupt drop, which was interpreted as the signature of two-dimensional antiferromagnetic (AFM) interactions.

In this paper, we have extensively investigated the transport and magnetic properties of $Ba_4Ru_3O_{10}$ by means of magnetic susceptibility, resistivity, thermoelectric power and Hall coefficient measurements, combined with neutron powder diffraction experiments. We show that $Tl_2Ru_2O_7$ and $La_4Ru_2O_{10}$ alike, the magnetic susceptibility drops at $T_N$ = 105K, precisely where the crossover between two semiconducting regimes is observed. Unlike the two latter compounds however, no structural transition is evidenced at $T_N$. Based on the observed magnetic scattering below $T_N$ on the neutron diffraction data, a model of three-dimensional antiferromagnetic ordering of the Ru(2) spins is proposed, in which Ru(1) at the centre of a trimer bears no ordered magnetic moment. This model is discussed in the framework of the strong spin-orbit coupling attributed to $Ru^{4+}$ ions.

**Experimental**

Polycrystalline samples of $Ba_4Ru_3O_{10}$ were synthesized by solid state reaction of $BaCO_3$ and $RuO_2$. Mixtures were first heated up to 1000°C in air for 24h, and then to 1200°C for 24h, with an intermediate grinding to avoid Ru evaporation. The resulting powder was then pelletized with a uniaxial press and sintered in air at 1400°C for 24h. The samples obtained following this procedure were checked by room temperature X-ray diffraction and found to be single phase and well crystallized $Ba_4Ru_3O_{10}$.

Neutron powder diffraction (NPD) versus temperature was performed on the G4.1 high-flux diffractometer ($\lambda$ = 2.425 Å) from 1.5 to 300K, and high resolution neutron diffractograms were recorded on the diffractometer 3T2 ($\lambda$ = 1.225 Å) at 10 K and 300 K. Both diffractometers are located at LLB-Orphée (CEA-Saclay, France). Rietveld refinements [13] and determination of the magnetic structure were performed with programs of the FullProf suite [14].

Magnetization as a function of temperature was recorded with a *dc*-SQUID magnetometer in zero field cooled (ZFC) and field cooled (FC) modes. Transport measurements were carried out using a four-terminal method with electric current $I$ = 1 mA. Thermoelectric power ($S$) was determined by applying a gradient of 0.5 K.cm$^{-1}$ to the sample and by measuring the resulting voltage. The Hall coefficient ($R_H$) was measured between 300K and 4.2 K with a Physical Property Measurement System (PPMS) from Quantum Design. To eliminate the unwanted voltage arising from the misalignment of the voltage pads, the magnetic field ($H$) was swept from -5 T to 5 T with a 20 minute period at constant temperature. $R_H$ was calculated from the slope of the Hall resistance $R_{xy}$ through the relation $[R_{xy}(H)-R_{xy}(-H)]/(2H)$.

**Magnetic and transport properties**

Figure 2 shows the temperature dependence of the magnetic susceptibility ($\chi$), electrical resistivity ($\rho$) and thermoelectric power (S) from 300 K down to 2 K. $\chi$ increases when $T$ decreases and reaches a broad maximum around $T_{max} \approx$ 180 K. A subsequent drop followed by a minimum at $T \approx$ 40 K is observed before $\chi$ starts increasing again, in agreement with previous results [11]. ZFC and FC measurements (not shown) lead to similar curves. To determine more precisely the antiferromagnetic transition temperature $T_N$ in $Ba_4Ru_3O_{10}$,

d($\chi T$)/d$T$ is calculated in the inset of figure 2a : it shows a slowing down of the magnetic fluctuations below $T_N \approx 105$ K. At 2 K, magnetization $M$ is linear with $H$ up to 7 T (not shown), as expected in a system with strong antiferromagnetic interactions.

The antiferromagnetic transition at $T_N \sim 105$ K occurs simultaneously with a change in the transport regime (Figure 2b), and with a jump of $S(T)$ which reaches up to 275 µV.K$^{-1}$ (Figure 2c), and which characterizes an insulating state with a marked increase of the activation energy below $T_N$. The variation of $S$ with $T$ is linear for both $T < 30$ K ($S/T \approx 7$ µV.K$^{-2}$) and $T > 200$K ($S/T \approx 0.07$ µV.K$^{-2}$). In a first approach, for a two dimensional material, $S/T$ is inversely proportional to the carrier density [15], so that the change of slope between these two regimes characterizes a diminution by two orders of magnitude of the charge carrier density. This is in agreement with the observed jump in resistivity, with the ratio $\rho_{5K}/\rho_{100K}$ being close to $10^5$.

In the high temperature regime the resistivity can be modeled with a polaronic expression:

$$\rho(T) = \frac{T}{C}\exp\left(\frac{E_p}{k_B T}\right) \quad (1)$$

where C is a constant, $k_B$ is the Boltzmann constant and $E_p = E_g + E_w$ is the sum of the energy gap plus the characteristic hopping energy of polarons. This is represented in the inset of figure 2b by plotting ln[$\rho(T)/T$] as a function of 1/$T$. We can infer from the thermoelectric power value (18 µV.K$^{-1}$ at 300K) which is metal-like (inset of figure 2c) that $E_g = 0$, so that the transport properties at $T > T_N$ of Ba$_4$Ru$_3$O$_{10}$ are well described by small polarons with a jumping energy $E_w = E_p \equiv 600$K.

The Hall coefficient is plotted on figure 3. As expected from thermoelectric power measurement, $R_H$ also carries the signature of the transition. For $T > T_N$, $R_H$ is negative and linearly decreases from -4.8x10$^{-4}$ cm$^3$.C$^{-1}$ at 300 K down to -8.7x10$^{-4}$ cm$^3$.C$^{-1}$ at 120 K. Considering the positive sign of $S$, Ba$_4$Ru$_3$O$_{10}$ is a two-carrier conductor (holes ($p$) and electrons ($e$)) at high temperature. At the transition temperature, $R_H$ suddenly changes its sign and amplitude to reach a value of 2.7x10$^{-2}$ cm$^3$.C$^{-1}$ at 100 K. By neglecting the minority $e$ carriers, we can roughly estimate that the carrier number decreases by a factor of 30 at the crossover between the two regimes. For $T < T_N$, $R_H$ continues to increase up to a value of 8.3x10$^{-1}$ cm$^3$.C$^{-1}$ at 10 K, in qualitative agreement with the above thermoelectric power results. As in an insulator, only one kind of carriers can conduct at low temperature because minority carriers have a limited thermal activation, the carrier concentration ($p$) and the carrier mobility ($\mu_p$) can be estimated without any difficulties : at 10 K, using the relations $p = 1/(q.R_H)$ and $\mu_p = R_H/\rho$, we obtain $p \approx 1.2 \times 10^{19}$ cm$^{-3}$ and $\mu_p \approx 4.4 \times 10^{-4}$ cm$^2$.V$^{-1}$.s$^{-1}$. For an oxide, the value of $p$ is extremely low and corresponds to ~0.003 holes/f.u.

**Structural analysis**

Ba$_4$Ru$_3$O$_{10}$ has been first reported to crystallize in the monoclinic $P2_1/c$ space group by Dussarat *et al.* [11], but later on Carim *et al.* suggested a more symmetric orthorhombic space group *Cmca* [12]. Note that, as explained in details in Ref. 12, the X-ray powder diffraction patterns calculated for both space groups are difficult to distinguish, and the differences between the two structures are accordingly minute. However, based on anisotropic thermal parameters considerations, the monoclinic space group was considered less plausible. Agreement factors obtained here from Rietveld refinements of X-ray and high resolution neutron powder diffraction data are comparable for both space groups. As a result, in agreement with Ref. 12, Ba$_4$Ru$_3$O$_{10}$ is described in the following using the space group of the highest symmetry. The 300 K neutron powder diffraction pattern (3T2) can be indexed in the orthorhombic unit cell with the space group *Cmca* with $a = 5.782(1)$ Å, $b = 13.279(1)$ Å, and $c = 13.088(1)$ Å (Fig. 4). No symmetry lowering is observed in the 300K-10K range. The thermal variation of the unit cell parameters (from G4.1 data) as a function of temperature is illustrated in Figure 5. It shows a regular and fairly isotropic lattice contraction, with no noticeable accident at $T_N$. Table 1 shows the unit cell parameters and atomic positions at 300 K and 10 K and table 2 shows the main atomic bond distances and bond angles: within the experimental resolution, there is no significant change of the Ru(1) or Ru(2) environments ; the intra-trimer Ru(1)-Ru(2) distance, as well as the Ru(2)-O(1)-Ru(2) bond angle between trimers, are also almost constant.

Careful reading of the data also shows extra scattering intensity on one Bragg peak when cooling the sample under $T_N$. On figure 6, a detailed view of the corresponding angular area ($2\theta \approx 21.5°$, *i.e.* $d \approx 6.53$ Å) is displayed together with the integrated intensity of the peak (inset of Fig. 6): it shows clearly that the scattering intensity is constant down to 120 K, and that it increases progressively as the temperature is lowered below $T_N$, so that it can reasonably be stated that it is of magnetic origin. The only magnetic component thus being seen on a crystalline Bragg reflection (0 0 2), the magnetic propagation vector is simply **k** =

(0 0 0). With the *Cmca* space group and **k** = (0 0 0), there are 4 one-dimensional irreducible representations associated with site 4*a* (Ru(1)) $\Gamma_{mag} = 1\Gamma_1 \oplus 2\Gamma_3 \oplus 2\Gamma_5 \oplus 1\Gamma_7$, and 8 with site 8*f* (Ru(2)), $\Gamma_{mag} = 1\Gamma_1 \oplus 2\Gamma_2 \oplus 2\Gamma_3 \oplus 1\Gamma_4 \oplus 2\Gamma_5 \oplus 1\Gamma_6 \oplus 1\Gamma_7 \oplus 2\Gamma_8$. Bearing in mind that we have only one magnetic peak, we will first consider the simple case of a second-order magnetic transition, with both atoms ordering with the same unique irreducible representation. To have an ordered magnetic moment on both sites, the magnetic structure should therefore follow representation $\Gamma_1$, $\Gamma_3$, $\Gamma_5$ or $\Gamma_7$. $\Gamma_7$ corresponds to a purely ferromagnetic arrangement of the moments and is not compatible with the susceptibility data. The three other magnetic configurations do not reproduce the intensity on the (0 0 2) peak properly. The remaining possible configurations allowed by symmetry ($\Gamma_2$, $\Gamma_4$, $\Gamma_6$ or $\Gamma_8$) give an ordered magnetic moment on the Ru(2) site only. Amongst them, $\Gamma_6$ and $\Gamma_8$ lead to similarly satisfying models. They correspond to an antiferromagnetic configuration with moments on the Ru(2) sites aligned along *a* or *b*, respectively (an antiferromagnetic component along *c* is allowed with $\Gamma_8$ but has not been refined considering the lack of magnetic reflections). Though it is difficult to decide between $\Gamma_6$ and $\Gamma_8$, we propose that the magnetic moment be aligned along the longest Ru-O distance of the RuO$_6$ octahedra, that is, along *b*, which therefore corresponds to the $\Gamma_8$ representation. The basis vectors of $\Gamma_8$ are listed in Table 3, and the corresponding magnetic arrangement is illustrated on Figure 7. It corresponds to antiferromagnetic zig-zag chains running along *a*, with the moments perpendicular to the plane of the chain. From experimental data, the magnetic moment on the Ru(2) site is estimated to be of the order of 1 $\mu_B$, which is half the value expected for Ru$^{4+}$ (S = 1) cations. There is no ordered moment on the Ru(1) site in this model, and the molecular field on the Ru(1) site is actually zero. It cannot completely be ruled out that Ru(1) and Ru(2) spin components could belong to different representations, which would be the sign of either a decoupling (spins along *a* or *c* for example) of the two species, or of a magnetic coupling of a lower symmetry indicative of higher order interactions. Within the accuracy of the experimental data, it seems however that antiferromagnetic ordering on the Ru(1) site (constraining the same moment on both Ru(1) and Ru(2) sites) is to be precluded, as it leads to new magnetic reflections. A small ferromagnetic component cannot be completely ruled out, however.

**Discussion**

One paradox in $Ba_4Ru_3O_{10}$, is the concomitant occurrence at $T > T_N$ of a Curie-like behavior together with a metallic-like thermoelectric power, weak Hall coefficient, and reasonably low electrical resistivity on the other hand. This is also known in the most metallic Ru-based oxides such as $SrRuO_3$ and $CaRuO_3$ [16]. The complicated bands structure of Ru-oxides can involve more than one kind of charge carriers [17, 18], as was experimentally observed in $SrRuO_3$, $CaRuO_3$ and $BaRuO_3$ [16, 19]. $Ba_4Ru_3O_{10}$ should be considered in this picture, because of the opposite signs of $R_H$ and $S$ at $T > T_N$. Charge carriers are differentiated by their sign (electrons or holes), concentration and mobility. Localized (itinerant) charge carriers have a strong (weak) contribution to the magnetic susceptibility but a weaker (stronger) contribution to the electrical conductivity. Therefore, a more quantitative description of the high temperature properties in $Ba_4Ru_3O_{10}$ would require a good knowledge of its band picture.

Above $T_N$, susceptibility data show that there is a clear contribution to the paramagnetic susceptibility of the three $Ru^{4+}$ ions. From an analysis of the high temperature susceptibility data of *Dussarrat et al.* [11] we could extract $\theta \approx -580$ K and $\mu_{eff} = 2.83$ $\mu_B$/Ru, in good accordance with paramagnetic $Ru^{4+}$ (S = 1) species. Below $T_N$, our simple magnetic model suggests that one third of the $Ru^{4+}$ spins are not ordered. We can first assume that they stay in a paramagnetic state : at $T < 50$ K, the upturn of the susceptibility can be due either to an impurity or to intrinsic paramagnetic spins. Fitting the low temperature susceptibility data with a Curie-Weiss law leads, however, to a very weak effective moment, that would correspond to less than 1% of paramagnetic S = 1 spins. In this context, it seems that magnetic moments on Ru(1) sites cannot stay in the paramagnetic state at low temperature. We can also suppose that the spins on the Ru(1) sites are randomly frozen at low temperature, but in this case one could expect a significant difference between zfc and fc susceptibility measurements, which is not observed. Our third hypothesis is based on the fact that, as a member of the second series of transition elements, the magnetic behavior of $Ru^{4+}$ is dominated by spin-orbit interaction effects at low temperature. In octahedral symmetry, the ground term of $Ru^{4+}$ is the orbital triplet $^3T_{1g}$ which splits into three sub-levels with energies $E = -2\lambda$ ($J = 0$), $E' = -\lambda$ ($J = 1$), and $E'' = \lambda$ ($J = 2$). Thus at low temperature only the $J = 0$ fundamental state would be occupied, and in this model $Ba_4Ru_3O_{10}$ should not show any long range order. To go further, it is useful to compare $Ba_4Ru_3O_{10}$ with 9R-$BaRuO_3$ in which basic units are also $Ru_3O_{12}$ trimers [20]. The main difference resides in the connections between these units. In 9R-$BaRuO_3$, each trimer is connected to six others trimers (three on both sides) through identical Ru(2)-O-Ru(2) 180° bonds, resulting in a three dimensional network.

Trimers have a ternary symmetry along *c*-axis which causes a trigonal distortion of Ru(1)O$_6$ and a trigonal-like distortion of Ru(2)O$_6$ [21]. According to Mössbauer spectrum, these distortions are not strong enough to remove the orbital degeneracy [22]. As a consequence, all Ru$^{4+}$ cations adopt the *J* = 0 fundamental state and 9R-BaRuO$_3$ do not order down to the lowest temperature [23]. This argument can also be applied to 4H-BaRuO$_3$ with a three dimensional framework of Ru$_2$O$_9$ dimers in which long range order is neither established [23, 24]. In comparison, trimers in Ba$_4$Ru$_3$O$_{10}$ are connected to four neighbors (two at both side) and one oxygen element, denoted as O(4), does not participate in any connection. This configuration breaks the third order axis of the trimer and allows a further distortion of Ru(1)O$_6$ and Ru(2)O$_6$. If the distortion index is defined as :

$$\sigma = \sqrt{\sum_{i=1}^{6}\left(1-\frac{R_i}{\frac{1}{N}\sum_{i=1}^{N}R_i}\right)^2} \qquad (2)$$

values of 0.009 (0.018) and 0.078 (0.075) are found for Ru(1)O$_6$ and Ru(2)O$_6$ at 300 K (10 K). This indicates that in Ba$_4$Ru$_3$O$_{10}$, Ru(1) environment is not strongly perturbed while Ru(2)-O bonds are highly modified by the lost of the ternary axis. As a consequence the crystal field is different on both sites, and it might lift the degeneracy of the *t$_{2g}$* orbitals thus destabilizing the *J* = 0 level of Ru$^{4+}$(2). Experimental spectroscopic data are necessary here to confirm that the degeneracy of the *t$_{2g}$* orbitals is lifted for Ru(2) and not for Ru(1). It is also possible that, even with a destabilized spin-orbit coupling of Ru(2), some angular orbital momentum could still contribute to the total effective momentum, *i.e. J* = |L-S| < 1. This would be in qualitative agreement with the reduced momentum of 1 μ$_B$ deduced from the refinement of the neutron diffraction data.

    Neutron diffraction clearly evidences a long range order and the magnetic coupling between the antiferromagnetic zig-zag chains must be supported by a reasonably strong exchange interaction along the trimers in relation with the high value of *T$_N$*. Taking into account an isolated trimeric (or dimeric) unit, *Drillon et al.* have developed the following expression for the exchange Hamiltonian that can be applied to 9R-BaRuO$_3$ (or 4R-BaRuO$_3$) :

$$H_{ex} = -J_{ex}/2 \sum_{i,j} \vec{L}_i \cdot \vec{L}_j (1 + \vec{L}_i \cdot \vec{L}_j)(1 + \vec{S}_i \cdot \vec{S}_j) \qquad (3)$$

with $J_{ex}$ being the exchange coupling between $d$ orbitals pointing towards the same edge of the face shared by two adjacent sites [25]. Their model gives a non-zero, weakly temperature dependent susceptibility, in good accordance with experimental data [23]. In particular it can reproduce the broad maximum of $\chi(T)$ if the exchange parameter $J_{ex}$ is not too weak in comparison with the spin-orbit coupling parameter $\lambda$. This suggests that, in the case of $Ba_4Ru_3O_{10}$, even if $Ru^{4+}(1)$ exhibits a non-magnetic behavior ($J = 0$), the exchange interaction between $Ru^{4+}(1)$ and $Ru^{4+}(2)$ $t_{2g}$ orbitals is strong enough to allow AFM coupling between zig-zag chains along the trimers.

In the single-layered $Ca_{2-x}Sr_xRuO_4$ ($x \leq 0.2$) system, an orbital selective Mott transition has been observed [26]. In this compound, the metal-insulator transition is accompanied by a paramagnetic-antiferromagnetic transition and is also associated with a structural transition inducing significant distortions of the $RuO_6$ octahedra [27]. All Ru-O bond distances exhibit variations of the order of 3%, a variation that is significant enough to be coupled with a change in the orbital occupation. As already mentioned above, $Ba_4Ru_3O_{10}$ does not experience any crystallographic transition. The main change observed is an increase of only 0.7% of the Ru(1)-O(2) bond distance, which should not be enough to significantly affect the occupation of the $t_{2g}$ orbitals. Nonetheless, Ru(1) and Ru(2) having different environments, the possibility that they exhibit different orbital occupations has to be considered. In a rather naïve picture, electrons on the Ru(2) sites seem to be localized, with a local moment, like in $Ca_2RuO_4$, while electrons on the Ru(1) site are likely to be more or less itinerant, like in $Sr_2RuO_4$. In fact, orbital occupation depends on a subtle balance of the local coordination of the $RuO_6$ octahedra [28], and the localized and itinerant duality of the Ru trimer structure in $Ba_4Ru_3O_{10}$ could be the way to understand its anomalous physical properties. "Molecular orbitals" of the Ru trimer may indeed play an important role, considering in particular that a similar pseudogap, with a $J = 0$ ground state, opens in the three dimensional trimer lattice of ruthenate 9R-$BaRuO_3$ [19]. However, a detailed band calculation of $Ba_4Ru_3O_{10}$ is required at this stage to go further in the understanding of this material. It is also interesting to note that, although $Ba_4Ru_3O_{10}$ shows a similar magnetic susceptibility to the Haldane spin-gap system $Tl_2Ru_2O_7$ [5], its physical properties are interpreted very differently. In $Tl_2Ru_2O_7$, the spontaneous orbital ordering associated with the crystallographic distortion leads to $S = 1$ antiferromagnetic chains, but also isolates these chains from each other because of a weak interchain coupling ($J_{inter}$ = 21 K), in comparison with the strong intrachain coupling ($J_{intra}$ = 268 K). In contrast, there is no evidence of low dimensionality in $Ba_4Ru_3O_{10}$ : three dimensional long-range magnetic ordering at low temperature evidences a strong antiferromagnetic coupling between zig-zag chains, that is mediated through the trimer units, emphasizing again the importance of the trimer role in the compound.

**Conclusion**

$Ba_4Ru_3O_{10}$ exhibits below $T_N$ = 105K a long-range antiferromagnetic phase with strongly localized carriers. The observation of magnetic scattering in neutron diffraction allowed us to propose a magnetic configuration in which spins on the Ru(2) sites are coupled antiferromagnetically to form zig-zag chains, while the Ru(1) site has no ordered magnetic moment, probably because of a $J$ = 0 fundamental state coming from a strong spin-orbit coupling that is not lifted by the local crystal field. Theoretical calculations are needed at this stage to understand the interplay between transport properties and orbital and spin momenta.

**Acknowledgments**

This work was supported by the Academic Frontier Project from MEXT.

**Figure captions**

Figure 1: (Color online) Crystallographic structure of $Ba_4Ru_3O_{10}$. Large yellow spheres represent Ba atoms, blue and white polyhedrons are $RuO_6$ octahedrons with Ru atoms at the center and O atoms (small red spheres) at the extremities. On the left-hand side, trimeric units form a checkered structure while on the right-hand side we can see the stacking of these layers along the *b*-axis.

Figure 2: (Color online) (a) Magnetic susceptibility as a function of temperature. The red curve shows the approximation to the modified Curie-Weiss law: $\chi = \chi_0 + C/(T-\theta)$. Inset shows $d(\chi.T)/dT$ in the region of the transition. (b) Resistivity as a function of temperature. Inset shows $\ln[\rho(T)/T]$ as a function of $1/T$. The red line corresponds to the approximation to equation (1). (c) Thermoelectric power as a function of temperature. Inset is an enlargement of $S(T)$ in the high temperature region. The red curve is a linear fit to the data.

Figure 3: Hall coefficient of $Ba_4Ru_3O_{10}$. Note that on the upper panel (a) *y*-axis follows a logarithmic scale with positive values while on the lower panel (b) *y*-axis follows a linear scale with negative values. Inset: mobility as a function of temperature.

Figure 4: (Color online) Room temperature neutron diffraction pattern of $Ba_4Ru_3O_{10}$ on the high resolution 3T2 diffractometer ($\lambda=1.225$ Å). From top to bottom, (i) red dots and black curve show the observed and calculated diffraction patterns respectively, (ii) green sticks indicate Bragg (*hkl*) positions, (iii) and blue line is the difference between the diffracted intensity and the refinement profile.

Figure 5: (Color online) Cell volume and parameters of $Ba_4Ru_3O_{10}$ as a function of temperature (from G4.1 data). Lines are guide to the eye.

Figure 6: (Color online) Detailed view of the diffraction patterns (G4.1 data) in the region $20° < 2\theta < 23°$. Inset shows the integrated intensity of the (002) Bragg reflection as a function of temperature.

Figure 7: (Color online) (a) Proposed magnetic arrangement of $Ba_4Ru_3O_{10}$ with the $\Gamma_8$ representation. (b) A corrugated layer projected in the *ac*-plane. Ru(2)-O-Ru(2) AFM zig-zag chains are represented in red. (c) Local magnetic coupling between Ru(2) moments within a trimer. (outlined in grey in (b)).

Table 1: Cell parameters, atomic positions and refinement agreement factors of $Ba_4Ru_3O_{10}$ at 300 K and 10 K.

Table 2: Selected bond lengths (Å) and bond angles.

Table 3: Basis functions for axial vectors associated with irreducible representation $\Gamma_8$ for Wyckoff site 8*f* of the *Cmca* space group, with the propagation vector **k** = (0 0 0).

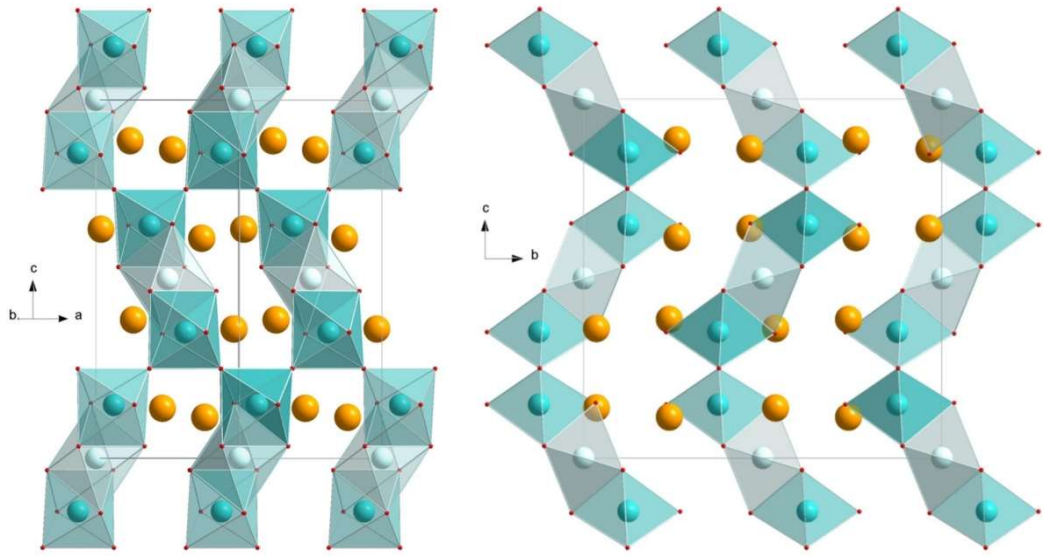

Figure 1

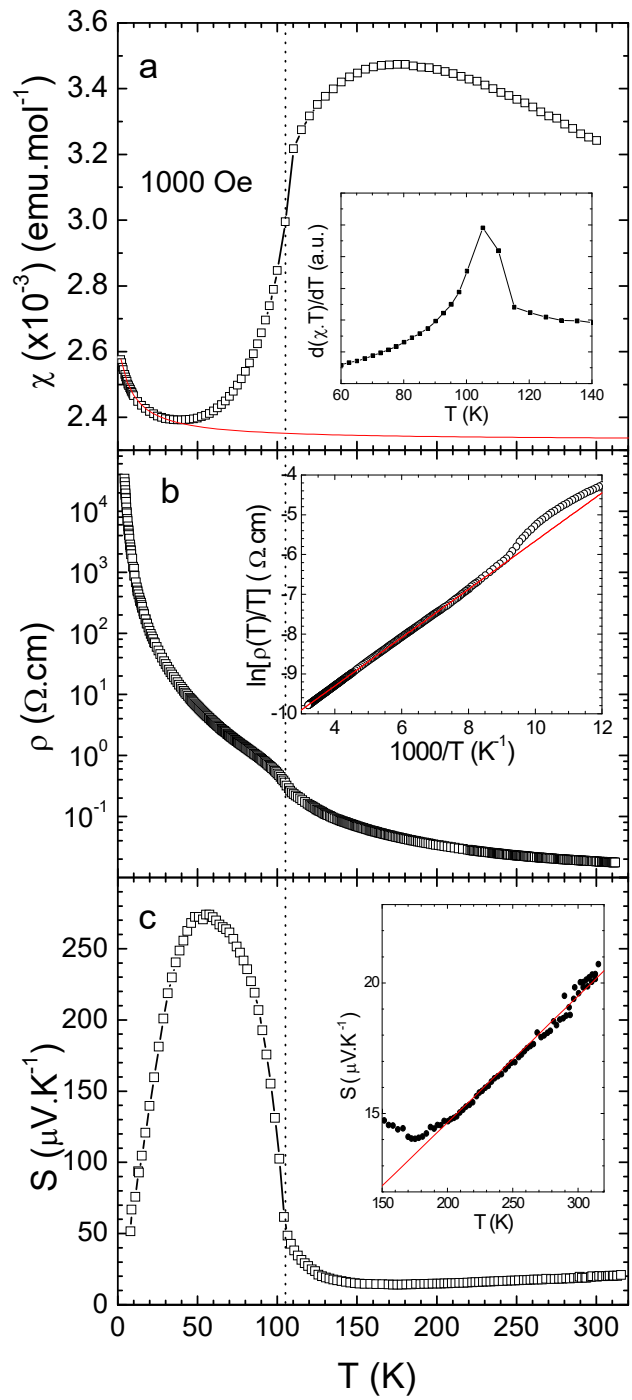

Figure 2

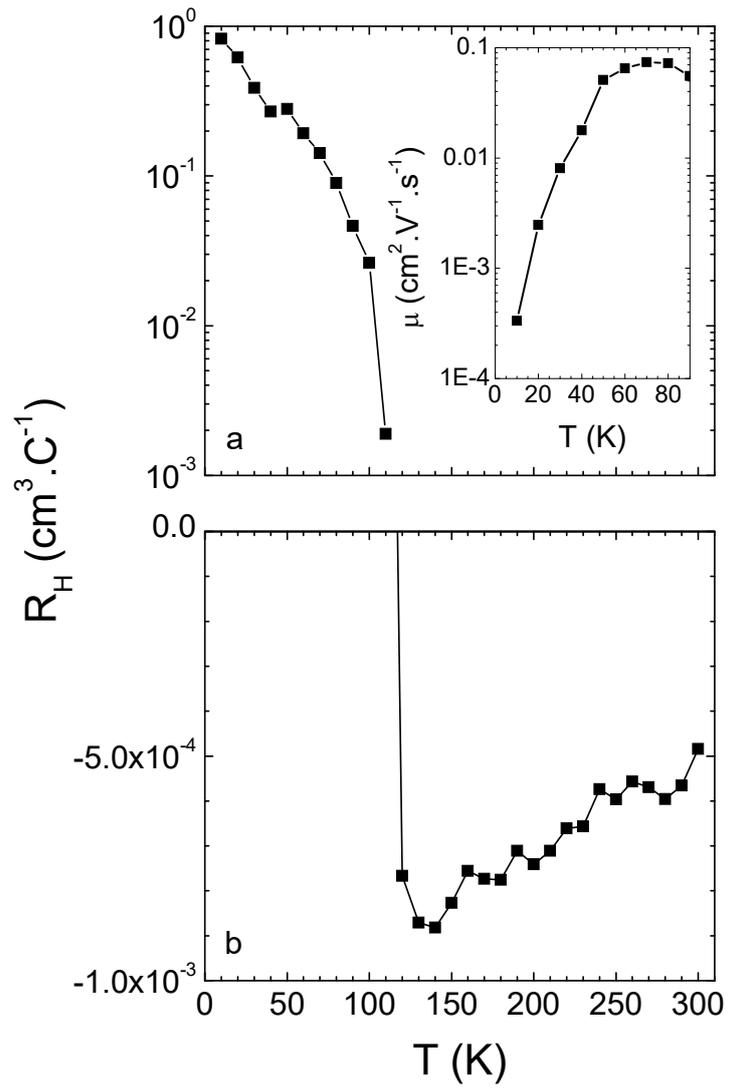

Figure 3

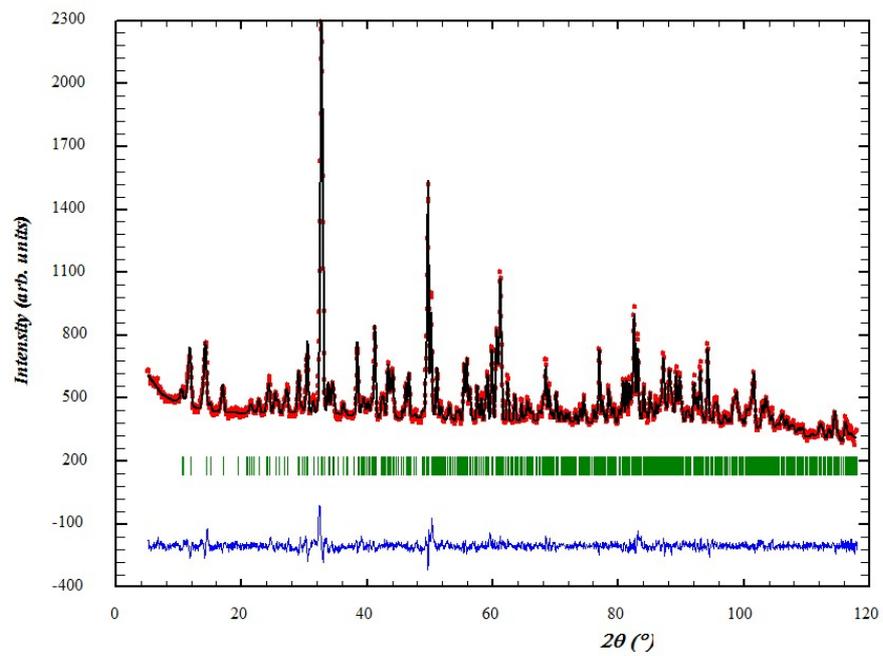

Figure 4

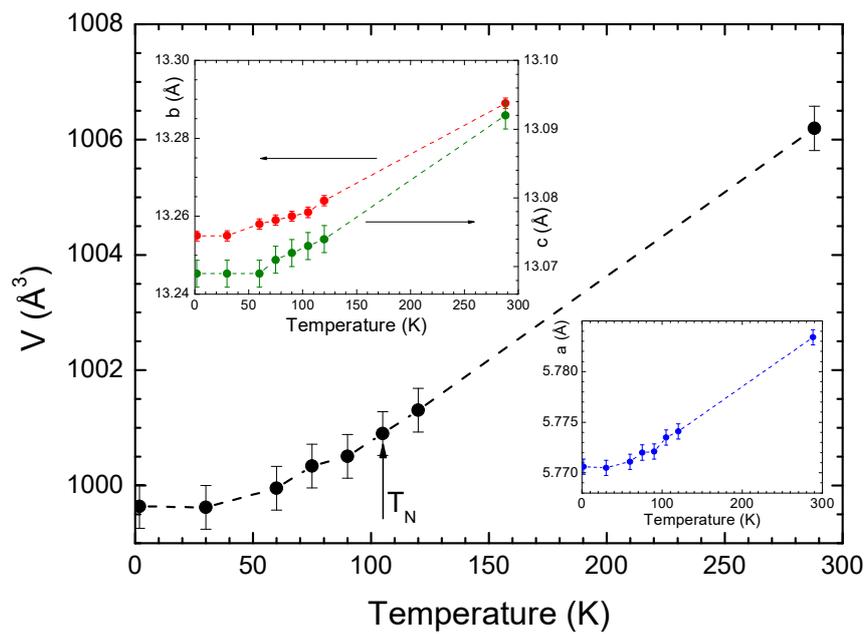

Figure 5

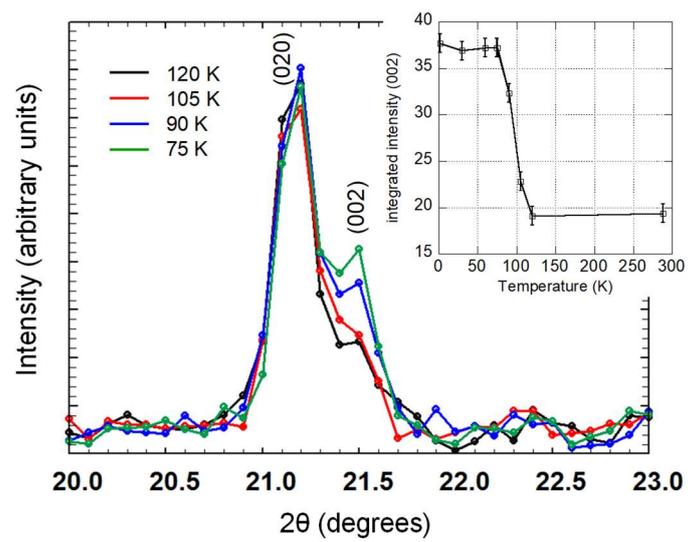

Figure 6

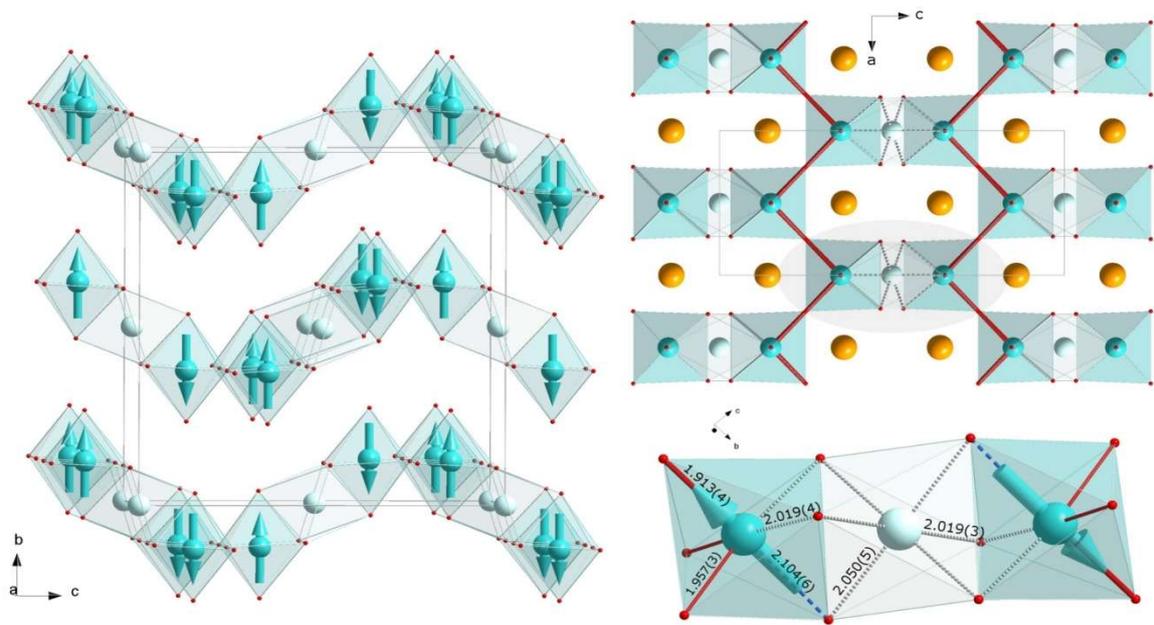

Figure 7

|  | Site |  | 300K | 10K |
|---|---|---|---|---|
| $a$ (Å) |  |  | 5.782(1) | 5.767(1) |
| $b$ (Å) |  |  | 13.279(1) | 13.245(2) |
| $c$ (Å) |  |  | 13.088(1) | 13.064(2) |
| $V$ (Å$^3$) |  |  | 1004.8 | 997.9 |
| Ba(1) (0,$y$,$z$) | 8$f$ | $y$ | 0.2403(5) | 0.2396(4) |
|  |  | $z$ | 0.1132(4) | 0.1141(3) |
| Ba(2) (0,$y$,$z$) | 8$f$ | $y$ | 0.5359(4) | 0.5358(3) |
|  |  | $z$ | 0.1398(4) | 0.1391(4) |
| Ru(1) (0,0,0) | 4$a$ |  |  |  |
| Ru(2) (0,$y$,$z$) | 8$f$ | $y$ | 0.8752(3) | 0.8753(3) |
|  |  | $z$ | 0.1484(3) | 0.1487(3) |
| O(1) (0.25,$y$,0.25) | 8$e$ | $y$ | 0.3776(4) | 0.3775(3) |
| O(2) (0,$y$,$z$) | 8$f$ | $y$ | 0.0344(4) | 0.0341(3) |
|  |  | $z$ | 0.1516(4) | 0.1531(4) |
| O(3) ($x$,$y$,$z$) | 16$g$ | $x$ | 0.2692(5) | 0.2693(5) |
|  |  | $y$ | 0.3906(3) | 0.3902(2) |
|  |  | $z$ | 0.0332(2) | 0.0334(2) |
| O(4) (0,$y$,$z$) | 8$f$ | $y$ | 0.7310(4) | 0.7308(3) |
|  |  | $z$ | 0.1499(4) | 0.1492(4) |
| R$p$ (%) |  |  | 14.7 | 12.8 |
| R$_{wp}$ (%) |  |  | 14.5 | 12.8 |
| R$_{exp}$(%) |  |  | 9.04 | 7.96 |
| $\chi^2$ |  |  | 2.60 | 2.62 |
| Bragg R-Factor |  |  | 6.92 | 6.68 |

Table 1

|  | 300K | 10K |
|---|---|---|
| Ru(1)-O(2) | 2.036(5) x2 | 2.050(5) x2 |
| Ru(1)-O(3) | 2.020(3) x4 | 2.019(3) x4 |
| Mean Ru(1)-O | 2.025(2) | 2.029(4) |
| Ru(2)-O(1) | 1.965(3) x2 | 1.957(3) x2 |
| Ru(2)-O(2) | 2.115(6) | 2.104(6) |
| Ru(2)-O(3) | 2.024(4) x2 | 2.019(4) x2 |
| Ru(2)-O(4) | 1.914(7) | 1.913(4) |
| Mean Ru(2)-O | 2.001(2) | 1.995(2) |
| Ru(1)-Ru(2) | 2.553 | 2.550 |
| Ru(2)-Ru(2) | 3.930 | 3.914 |
| Ru(2)-O(1)-Ru(2) | 178.1(2)° | 178.3(2)° |

Table 2

| $\Gamma_8$ | (x y z) | (-x -y+½ z+½) | (-x y+½ -z+½) | (x -y -z) |
|---|---|---|---|---|
| $\psi_1$ | 0 1 0 | 0 1 0 | 0 $\bar{1}$ 0 | 0 $\bar{1}$ 0 |
| $\psi_2$ | 0 0 1 | 0 0 $\bar{1}$ | 0 0 1 | 0 0 $\bar{1}$ |

Table 3